# Pinning in nonmagnetic borocarbides

A. N. Zholobenko, G. P. Mikitik, and V. D. Fil[a)]

*B. I. Verkin Institute for Low Temperature Physics and Engineering of the National Academy of Sciences of Ukraine, pr. Lenina 47, Kharkov 61103, Ukraine*

D. V. Fil

*Institute of Single Crystals of the National Academy of Sciences of Ukraine, pr. Lenina 60, Kharkov 61001, Ukraine*

J. D. Kim and E. M. Choi

*Pohang University of Science and Technology, Pohang 794-784, Republic of Korea*

S. I. Lee

*Korea Basic Science Institute, Daejeon 305-333, Republic of Korea*

The field dependences of the Labush parameter in nonmagnetic borocarbides are measured by a method that does not require achieving a critical state. The expected values of the critical current are estimated. The values obtained are two orders of magnitude greater than the results of "direct" measurements performed on the basis of transport (magnetic) experiments. A giant peak effect, which the collective pinning model describes quantitatively well, is observed in the field dependences of the Labush parameter in Y-containing borocarbides.

The dynamics of vortex matter in type-II superconductors is determined by the ratios of the elasticity of the fluxoid lattice, the viscosity, and the pinning. The study of such dynamics still attracts a great deal of attention even though it has been pursued for the past 40 years. This refers to, first and foremost, the characteristics of pinning because of the practical importance of understanding its physical nature in detail.

At the present time the intensity of pinning is ordinarily characterized, primarily, by the current density $j_c$, corresponding to the achievement of a critical state, i.e. a transition from a regime of dissipation-free current flow to a regime of free motion of vortices. Pinning can also be characterized by the "spring" Labush parameter $\alpha_L = d^2W_p/dx^2$, which determines the average curvature of the pinning potential $W_p$. A transition into the critical state corresponds to the Lorentz force being equal to the effective pinning force:

$$\alpha_L \xi \approx \frac{1}{c} j_c B, \qquad (1)$$

where $\xi$ is the coherence length and $B$ is the induction in the sample.

Using the relation (1) it is easy to estimate the value of $j_c$ to be expected for known $a_L$ and compare it with "direct" measurements.

In the present communication the results of measurements of $\alpha_L$ in nonmagnetic borocarbides ($YNi_2B_2C$, $Y_{0.95}Tb_{0.05}Ni_2B_2C$, and $LuNi_2B_2C$), obtained by a method that does not require reaching a critical state, are presented. It was found that our estimates of $j_c$ are two orders of magnitude higher than the critical currents obtained in transport or magnetic measurements. In addition, a giant peak effect was found in the field dependences of $\alpha_L$ for Y-containing samples. It can be described quantitatively well on the basis of a collective pinning model.[1] In leutecium borocarbide, pinning on defects, whose range is greater than the core size, is also found to be substantial.

The method is based on analysis of the amplitude and phase of the electromagnetic field emitted from a conducting medium under the action of a transverse sound wave propagating along the magnetic field $H$. For a uniform half-space and an elastic free interface the induction (Hall) component of the field is described by the simple expression:[2,3] [1)]

$$E_{ind} = \frac{[\dot{u}B]}{c} \cdot \frac{k^2}{q^2 + k^2} \equiv \frac{[\dot{u}B]}{c} \cdot X(B), \qquad (2)$$

where $u$ is the amplitude of the displacements in the elastic wave at the interface, $q$ is the wave number of the sound, and $k$ is the skin wave number of the experimental medium. In the normal state $k^2 = k_n^2 = (4\pi i\omega\sigma_0)/c^2$ and $\sigma_0 = (ne^2\tau)/m$ is the static conductivity.

In the mixed phase $k^2 = k_m^2 = 4\pi(i\omega\eta + \alpha_L)/B^2$, where $\eta$ and $\alpha_L$ are, respectively, the viscosity and Labush parameter per unit volume.

Since $\eta$ is approximately proportional to $B$ ($i\omega\eta \approx k_n^2 B H_{C2}/4\pi$, the Bardeen-Stephen relation[4]), in sufficiently weak field $|k_m^2| \gg q^2$ and $X(B)$ is close in amplitude to 1 and its phase is close to zero.

For $\kappa \gg 1$ ($\kappa$ is the Ginzburg-Landau parameter), in the actual region of the fields we need not distinguish between the induction in the sample and the applied field. Normalizing the measured value of $E/H$ so that for $H \sim (5-10)H_{C1}$ its modulus is close to 1, and taking the phase $\Phi$ in these fields as the point of reference, we obtain the field dependence of the complex quantity $X(H)$.

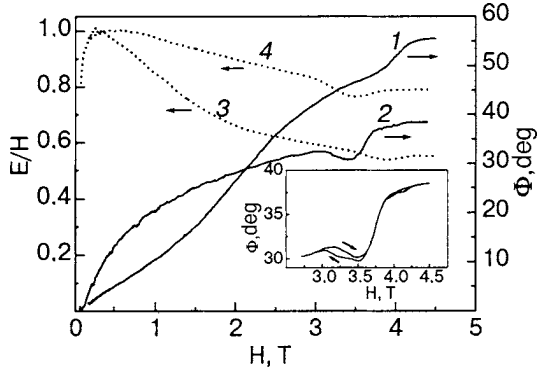

FIG. 1. Typical field dependence of the amplitude and phase of $X(H)$. The scale on the left-hand side is normalized as explained in the text: $\Phi$ (1), (2); $E/H$ (3), (4); LuNi$_2$B$_2$C ($T=6$ K) (1), (3); Y$_{0.95}$Tb$_{0.05}$Ni$_2$B$_2$C ($T=1.7$ K) (2), (4). Inset—hysteresis of the phase near the peak effect in Y$_{0.95}$Tb$_{0.05}$Ni$_2$B$_2$C ($T=1.7$ K).

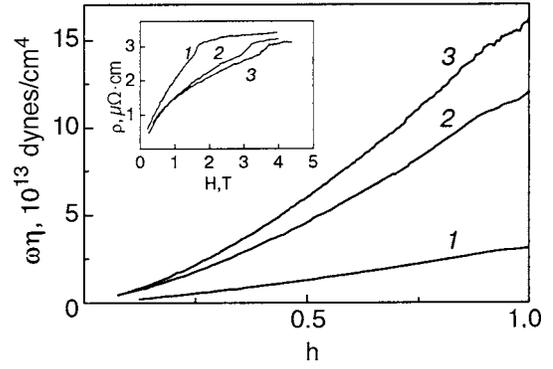

FIG. 2. Field dependences of $\omega\eta$ in Y$_{0.95}$Tb$_{0.05}$Ni$_2$B$_2$C: $T=8$ K, $H_{C2}=1.7$ T (1); $T=4$ K, $H_{C2}=3.3$ T (2); $T=1.7$ K, $H_{C2}=3.8$ T (3). Inset—expected variation of the resistivity. The labeling is the same as in the main figure.

The values of the viscosity and the Labush parameter are determined by the relations

$$\alpha_L = \mathrm{Re}\left(\frac{X(H)}{1-X(H)}\right) \cdot \frac{q^2 H^2}{4\pi}, \tag{3}$$

$$\omega\eta = \mathrm{Im}\left(\frac{X(H)}{1-X(H)}\right) \cdot \frac{q^2 H^2}{4\pi}. \tag{4}$$

It should be remembered that the information obtained in this experiment refers to a thin ($\sim q^{-1}$) layer near the surface. If this layer is nonuniform, the simple relations (2)–(4) break down. This question is studied in detailed in Ref. 5. It is shown that the nonuniformity of $\sigma_0$ (the decrease of the conductivity of a layer near the surface) increases the phase angle, fixed in the normal state. However, the nonuniformity of pinning results in an apparent nonmonotonic variation of the parameter $\eta$, if the relation (4) is used to reconstruct $\eta$, right up to the appearance of nonphysical negative values of the viscosity.

We shall indicate a simple test for revealing at least the nonuniformity of $\sigma_0$: in the normal state the modulus and phase of $X(H)$ in a uniform material should be related as

$$|X(H)| = (1+\tan^2\Phi(H))^{-0.5}.$$

In the experiments described below this condition always holds to within $\sim 5\%$.

The samples were grown by the standard technology used for compounds of this class.[6] They were in the form of thin $\sim 0.5$ mm thick flakes with transverse size $\sim 3$ mm. The $C_4$ axis was always orthogonal to the plane of the platelet. A quite perfect face of natural growth was used as the emitting surface. The opposite face was polished to create a reliable acoustic contact with a germanium delay line, making it possible to separate the exciting and analyzed signals in time. The excitation frequencies were 54–55 MHz. The details of the measurement procedure are described in Ref. 5.

Examples of typical experimental dependences of the modulus and phase of $X(H)$ are presented in Fig. 1. All experimental samples, irrespective of composition, had close values of $k_n^2$ and, correspondingly, the residual resistivity $\rho_{\mathrm{res}} \sim 3$ $\mu\Omega\cdot$cm and London penetration depth $\lambda(0)$ $\sim 10^{-5}$ cm. The values of the velocities, required for these estimates, of the $C_{44}$ modes are presented in Ref. 7.

The field dependences of the viscosity, which also turned out to be similar for identical values of $H_{C2}$, were found, using Eq. (4), from data similar to those presented in Fig. 4. An example is presented in Fig. 2. Their behavior is close to that predicted by the Bardeen-Stephen phenomenological model,[4] although the deviations from a linear dependence are quite large. The inset in Fig. 2 shows the behavior of the resistivity expected in the resistive regime from the measured values of the viscosity. The field dependences $\eta(H)$ do not show sufficiently strong nonmonotonic behavior, which in accordance with the results of Ref. 5 indicates that there is no significant nonuniformity in the characteristics of pinning. We also call attention to the fact that the presence of substantial nonmonotonic behavior near $H_{C2}$ in the primary data (Fig. 1, curve 2) has essentially no effect on $\eta(H)$.

Examples of the field dependences of the Labush parameter are presented in Fig. 3. One notices first and foremost the giant peak effect in the Y-containing samples near $H_{C2}$. Traces of its existence are also seen in LuNi$_2$B$_2$C. We also note that the solution (3) gives negative values of $\alpha_L$ for Y$_{0.95}$Tb$_{0.05}$Ni$_2$B$_2$C in intermediate-value fields. We shall assume that for very weak pinning the accuracy of the procedure used to reconstruct the Labush parameter using Eq. (2) is inadequate because factors such as the nonuniformity of the near-surface layer or thermal fluctuations are neglected, and the result $\alpha_L<0$ is a kind of artifact.

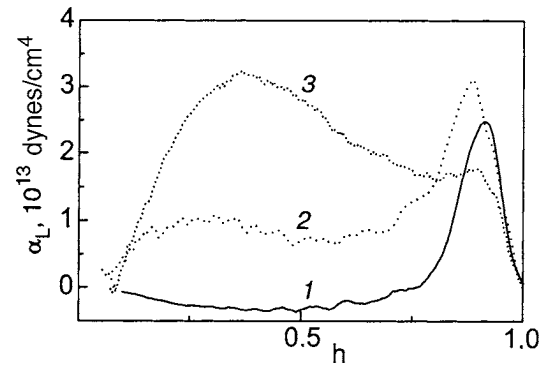

FIG. 3. Field dependences of $\alpha_L$: Y$_{0.95}$Tb$_{0.05}$Ni$_2$B$_2$C ($T=1.7$ K, $H_{C2}=3.8$ T) (1); YNi$_2$B$_2$C ($T=4$ K, $H_{C2}=3.8$ T) (2); LuNi$_2$B$_2$C ($T=6$ K, $H_{C2}=4.2$ T) (3).

Pippard has proposed a qualitative explanation for the nature of the peak effect.[8] The structure of the mixed phase is determined by the competition between the pinning and the intervortex interaction force. When the latter predominates the fluxoids form a nearly translationally ordered lattice and the details of the relief of the pinning potential are largely ignored. As $H_{C2}$ is approached the elasticity of the vortex lattice decreases more rapidly than the pinning intensity, which results in more efficient adjustment of the structure to the relief of the pinning potential and, correspondingly, to an increase of $\alpha_L$ and $j_c$.

The theory of collective pinning (CP) of vortex structures by point defects (i.e. defects whose range is shorter than the coherence length) made it possible to convert these qualitative considerations into a quantitative foundation.[1] In the CP theory a single free parameter characterizing the pinning force is introduced. It is convenient to take as this free parameter the dimensionless magnetic field $h_{SV} \equiv H_{SV}/H_{C2}$, determining the boundary of the so-called *bundle vortex pinning* (BVP) regime $h_{SV} < h < 1 - h_{SV}$, where $h = H/H_{C2}$ and the transverse size $R_c$ of Larkin's correlation region is greater than the vortex lattice parameter $a = \sqrt{\Phi_0/H}$ ($\Phi_0$ is the flux quantum). In the BVP regime $\alpha_L$ is determined from the condition that the pinning energy ($\sim \alpha_L \xi^2 V_c$) is equal to the elastic energy ($\sim C_{66}(\xi/R_c)^2 V_C$) of the vortex lattice in the correlation volume $V_C$. This gives

$$\alpha L \approx \frac{C_{66}}{R_c(H)^2}. \qquad (5)$$

The shear modulus is defined by the relation[9]

$$C_{66} \cong \frac{\Phi_0 H_{C2}}{(8\pi\lambda)^2} h(1-h)^2. \qquad (6)$$

The equation for finding $R_c$ (with $a \ll \lambda$) has the form

$$\left[\frac{h(1-h)}{h_{SV}(1-h_{SV})}\right]^{3/2} \approx 1 + 2\ln\frac{R_c}{a} + \frac{R_c}{\lambda}(1-h)^{1/2}. \qquad (7)$$

Actually, Eq. (7) is Eq. (4.17) from Ref. 10, where the possibility that $H$ approaches $H_{C2}$ is taken into account (see Eq. (8) in Ref. 11 and the accompanying explanation). The Labush parameter in the *single vortex pinning* (SVP) regime, in the lower region with respect to the magnetic field ($h < h_{SV}$), is linear in the magnetic field with the coefficient of proportionality determined from the condition of matching with Eq. (5). In the upper SVP region ($1 - h_{SV} < h < 1$) the estimate $\alpha_L \approx C_{66}(h)/\beta a^2(h)$ can be used, where $\beta$ is a correction factor close to 1, which also provides matching with Eq. (5) for $h = 1 - h_{SV}$.

The computed values of $\alpha_L(h)$, constructed for $H_{C2} = 4$ T and $\lambda = 10^{-5}$ cm, characteristic for the experimental samples, for various values of the parameter $h_{SV}$ are presented in Fig. 4. A remarkable property of these dependences is the single-valued relation between the form of the field dependence and the scale of the variations of $\alpha_L(h)$. In other words if $\alpha_L(h)$ is bell-shaped without a distinct peak effect, then the maximum value of $\alpha_L$ must be at the level $10^{15}$ dynes/cm$^4$. Conversely, if the peak effect is pronounced, the observed values of $\alpha_L(h)$ should not exceed $10^{13}$–$10^{14}$ dynes/cm$^4$. Turning to Fig. 3 we immediately conclude that the behavior of $\alpha_L(h)$ in Y-containing samples can be analyzed from the standpoint of the CP theory, which cannot be said of leutecium borocarbide.

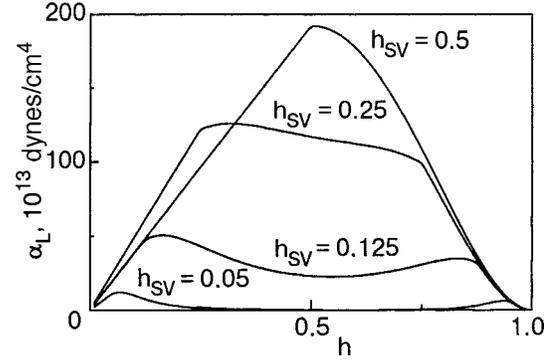

FIG. 4. Computed dependence $\alpha_L(h)$ ($H_{C2} = 4$ T, $\lambda = 10^{-5}$ cm) for various values of the parameter $h_{SV}$.

The CP theory predicts a nearly symmetric function $\alpha_L(h)$ with respect to $h = 0.5$, i.e. if a peak effect is observed near $H_{C2}$, then a rise of $\alpha_L(h)$ of the same type should also occur in weak fields. In our experiments we never observed such a dependence. However, we note that a low-field peak effect should occur in the field range where it is no longer possible to neglect the difference between $B$ and $H$ and Eqs. (3) and (4) become invalid. In addition, in these fields $X(H)$ is close to 1, and the relation (3), as already mentioned earlier, is very sensitive to the possible corrections which were neglected, including also to thermal fluctuations, which decrease the effective magnitude of pinning.[5]

Figure 5 demonstrates the "quality" of the description of the amplitude of the peak effect near $H_{C2}$ by the CP theory at various temperatures. This description is fully acceptable. The figure was constructed using the only adjustable parameter $h_{SV}(0) = 0.033$ and the temperature dependences $\lambda(t) = \lambda(0)(1-t^2)^{-1/2}$ and $h_{SV}(t) = h_{SV}(0)(1-t^2)^{1/2}$. The latter dependence corresponds to $\delta l$ pinning.[11]

The point of view that a transition into the peak-effect regime corresponds to a first-order phase transformation from a vortex lattice state into a disordered amorphous state is currently very popular (see Ref. 11 in the references cited there). It is shown in Ref. 11 that the position of this transition is correlated with the boundary of the upper region of the SVP regime. In other words, this point of view relates the peak effect not simply to a smooth transition from the BVP into the SVP regime but to a true phase transition. The inset in Fig. 1 demonstrates the hysteresis observed in our experiment. This hysteresis is characteristic for extended phase transformations, such as of the martensite type, and confirms this point of view.

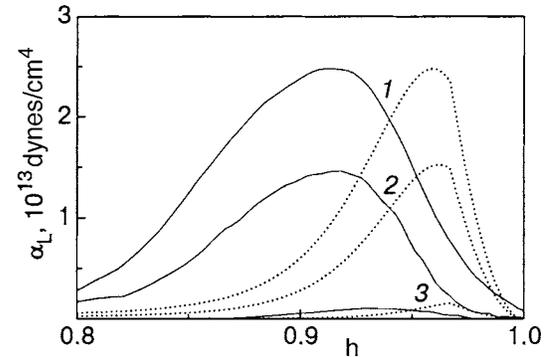

FIG. 5. Comparing the values of $\alpha_L$ measured in $Y_{0.95}Tb_{0.05}Ni_2B_2C$ (solid lines) and the computed values ($h_{SV} = 0.033$, points) under different conditions: $T = 1.7$ K, $H_{C2} = 3.8$ T (1); $T = 4$ K, $H_{C2} = 3.3$ T (2); $T = 8$ K, $H_{C2} = 1.7$ T (3).

As already mentioned above, the scale and form of $\alpha_L(h)$ in LuNi$_2$B$_2$C do not permit describing the field dependence of the Labush parameter on the basis of only the CP model. Since an indistinct peak for the same values of $h$ as in the Y-containing borocarbides is present on the right-hand wing of $\alpha_L(h)$, it is evident that weak pinning centers described by the CP model are also present in leutecium borocarbide. The main maximum at $h \sim 0.4$ is probably due to sparse but stronger pinning centers, whose range $r$ is greater than the coherence length.[1] The maximum value of $\alpha_L(h)$ is then described by the relation

$$\alpha_{L\,max} \approx \frac{nr^2}{\xi} \frac{H_{C2}^2}{\pi^3 \kappa^2},$$

where $n$ is the density of "strong" pinning centers. For $\kappa \sim 10$ and $r \sim 10^{-6}$ the approximate density $n \sim 10^{14}$ is indeed low.

It is of interest to compare our estimates of the critical currents with the values obtained in the transport (magnetic) measurements. We shall use Eq. (1) to calculate the critical current expected from the measured values of $\alpha_L$. For all experimental samples (with $H \sim 4$ T) we obtain $j_c \sim 10^4$ A/cm$^2$. In Ref. 12 the transport measurements in a LuNi$_2$B$_2$C crystal with $H \| c$ near $H_{C2}$ and $T=2.2$ K gave $j_c \sim 10^2$ A/cm$^2$. The same value is obtained in measurements of the irreversible magnetization in a YNi$_2$B$_2$C single crystal at $T=5$ K.[13] It is important to note that the single crystals studied in the works cited were grown, just as in the present investigation, using completely identical technologies. These samples should also have close pinning characteristics, and the two orders of magnitude difference in the measured values of the critical currents from our estimates is hardly due to the individual characteristics of concrete samples.

On the one hand, we note that in the works cited above it is most likely the current $j_t$ established in the sample over the measurement time in the thermally activated vortex flow regime is measured rather than $j_c$. The quantity $j_t$ can be estimated from the expression[10]

$$j_t \approx j_c \left[ 1 + \frac{\mu T}{U_c} \ln\left(1 + \frac{t}{t_0}\right) \right]^{-1/\mu}, \qquad (8)$$

where $t$ ($\sim 10^2 - 10^3$ s) is the characteristic measurement time in the experiments of Refs. 12 and 13, $t_0$ ($\sim 10^{-5}$ s) is a constant which depends on the conductivity and the size of sample,[10] and $U_c$ is an energy barrier which prevents free motion of a fluxoid. For the parameter $\mu$ at the boundary of the BVP and SVP regimes the estimate given by the CP theory is quite indefinite ($\mu \sim 1/7 - 5/2$). The pinning energy in the correlation volume $V_c$ ($U_c \approx \alpha_L \xi^2 V_c$) should be used as $U_c$. Near the peak effect $V_c \sim a^3$, $U_c \sim 1$ K, and $j_t/j_c \sim 10^{-2}$ is comparable to (8) for $\mu \sim 0.7 - 1$. Unfortunately, the lack of the required data in Refs. 12 and 13 (the current-voltage characteristics and the time evolution of the irreversible magnetization) precludes a quantitative check of the explanation presented.

The difference of the values of $j_t$ measured in Refs. 12 and 13 from our estimates of $j_c$ could have another reason, at least partially. The relation (1) in some sense should be understood as a condition for attaining the theoretical limit of elasticity in a vortex lattice. However, it is well known that in ordinary crystal lattices, as a rule, because of the presence of defects (dislocations), irreversible plastic deformations appear long before the moment of brittle fracture. A similar scenario can also be expected in vortex lattices with defects. We shall underscore the fact that it is precisely in borocarbides that vortex structures are characterized by a high density of defects,[12] which is due to the phase transformations, occurring in them, from a low-field hexagonal fluxoid lattice into a square high-field lattice. In such a case the estimates of the critical field on the basis of the CP theory must be modified, since they neglect the possibility of the existence of dislocations in the vortex lattice. Specifically, the relation (1), which presumes that the parameter $\alpha_L$ remains unchanged with small and large ($\sim \xi$) deformations, becomes invalid. In other words, fluxoid motion in defective vortex lattices with small supercriticality also start as plastic flow.

In closing, we shall formulate the basic results of this work. The field dependences of the Labush parameter were measured in single crystals of nonmagnetic borocarbides by a method that does not require reaching a regime of free flow of vortices. An estimate of the critical current based on these dependences gives values which are two orders of magnitude higher than the values obtained in transport (magnetic) measurements. This is the main result of the present work. In Y-containing samples, a giant peak effect was found in the field dependences of $\alpha_L$ near $H_{C2}$. Its magnitude (and temperature variations) are described well on the basis of the collective pinning model. In leutecium borocarbide, pinning on defects with range longer than the coherence length makes the main contribution to $\alpha_L$.

This work was performed under partial support by the CRDF Foundation (grant No. UP1-2566-KH-03) and INTAS (grant No. 03-51-3036).

---

[a)] E-mail: fil@ilt.kharkov.ua
[1)] Strictly speaking, a receiving antenna reacts to the high-frequency magnetic component $\widetilde{H}$. Near an interface the emitted field is a plane wave, so that $\widetilde{H} = \widetilde{E}$. The component $\widetilde{E}$, because of continuity, equals to within $\delta/\lambda_{EM} \sim 10^{-5}$ ($\delta$ is the skin depth and $\lambda_{EM}$ is the wavelength of the electromagnetic wave in vacuum) the field $E_{\text{ind}}$ generated by the elastic wave in the conductor.[2]

---